\documentclass[twocolumn,showpacs,preprintnumbers,amsmath,amssymb,prb]{revtex4}

\usepackage{graphicx}

\def\gz{\ifmmode{Z\hskip -4.8pt Z}
    \else{\hbox{$Z\hskip -4.8pt Z$}}\fi}

\newcommand{\be}{\begin{equation}}
\newcommand{\ee}{\end{equation}}
\newcommand{\bea}{\begin{eqnarray}}
\newcommand{\eea}{\end{eqnarray}}

\begin{document}

\title{Spin-charge separation in strongly interacting finite ladder rings}

\author{Juli\'an Rinc\'on, K.~Hallberg, and A.~A.~Aligia}
\affiliation{Centro At\'omico Bariloche and Instituto Balseiro, 
Comisi\'on Nacional de Energ\'ia At\'omica, 8400 Bariloche, Argentina}

\date{\today}

\begin{abstract}

 We study the
conductance through Aharonov-Bohm finite ladder
rings with strongly interacting electrons, modelled by the prototypical $t$-$J$
model. For a wide range of parameters we observe characteristic dips in the
conductance as a function of magnetic flux, predicted so far only in chains 
which are a signature of spin and charge separation. 
These 
results open the possibility of observing this peculiar many-body phenomenon
in anisotropic ladder systems and in real nanoscopic devices.

\end{abstract}


\maketitle

\section{Introduction}

The phenomenon of the fractionalization of an electron into its spin and charge
degrees of freedom was predicted theoretically for one dimensional (1D) strongly
interacting systems in the framework of the Luttinger Liquid (LL)  
theory.~\cite{Haldane,Schulz,kurt} Continuing progress in fabrication 
techniques, and the discovery of new materials of
quasi-1D electronic character, have led in the last decade to a variety of
experiments which seek evidence of spin charge separation (SCS) such as the
observation of non-universal power-law $I$-$V$ characteristics,~\cite{Voit1} the
search for characteristic dispersive features by angle-resolved photoemission
spectroscopy (ARPES),~\cite{Voit2} the violation of the Wiedemann-Franz law,~
\cite{Taillefer} and the analysis of spin and charge 
conductivities.~\cite{Voit2,Qimiao} 
Among the candidate materials to present SCS~\cite{yacoby} we can mention the
organic Bechgaard and Fabre salts, molybdenum bronzes and
chalcogenides,~\cite{Voit1} cuprate chain and ladder compounds~\cite{DagottoRice},
laterally confined two-dimensional electron gases, cleaved-edge overgrowth 
systems~\cite{2DEG} and also carbon nanotube systems.~\cite{egger,Byczuk}

From the theoretical point of view, several ways for detecting and visualizing SCS 
were proposed. Direct calculations of the real-time evolution of
electronic wave packets in Hubbard rings revealed that the spin and charge densities
dispersed with different velocities as an immediate consequence of 
SCS.~\cite{jagla1,uli} The analysis of the electronic transmission through 
Aharonov-Bohm (AB) rings~
\cite{Jagla,nos1,meden} described by a LL presented striking features characteristic of 
SCS, where the 
flux-dependence of the
transmission was found to show new structures appearing at fractional flux values 
in addition to the non-interacting flux quantum periodicity $\Phi_0 =hc/e$. 
In \cite{Jagla} these fractions were determined by the ratio between the spin and 
charge velocities $v_s/v_c$. In their interpretation 
the dips arise because transmission requires the separated spin and charge
degrees of freedom of an injected electron to recombine at the drain lead
after traveling through the ring a different number of turns in the presence
of the AB flux. However, recent results which go beyond the single pole  
approximation used there, suggest that this idea is too simple and claim that
the number of dips is not determined approximately by 
$v_{c}/v_{s}$ but by $v_{J}/v_{s}$, where $v_{J}$ is the current 
velocity.~\cite{meden}
The results, however, agree for small integer values of $p$ and $q$, for $v_s/v_c=p/q$.
Recent numerical calculations of the transmittance through finite AB rings 
described by the $t$-$J$ model show clear dips at the fluxes that correspond 
to the ratio $v_s/v_c$.~\cite{nos1} As we explain below, the discrepancy arises due to 
the finiteness of the system.

In spite of the clear indications of the existence of spin-charge separation in 1D 
interacting systems and its absence in three dimensions where the Fermi Liquid 
theory is valid, there is no final word for two dimensions. 
The non-Fermi-liquid
normal state properties of high temperature superconductors have led to
attempts to trace their origin in the possible realization of SCS in strongly
correlated electron systems in 2D.~\cite{Andersonbook}

In this paper we explore the possibility of the existence of SCS in ladders, 
as a first step towards two dimensions. 
We analyze the conductance through rings formed by two-leg ladder systems described by 
the $t$-$J$ model as a prototype of interacting systems. For certain 
parameters we find, indeed, clear dips at fractional values of the magnetic flux 
which we can interpret as fingerprints of charge and spin 
separation due to the difference in the charge and spin velocities.

\section{The model}

Our model Hamiltonian reads
$H  = H_{{\rm leads}} + H_{{\rm link}} + H_{{\rm ring}}$,
(Fig.~1), where
$H_{{\rm leads}}$ describes free electrons in the left and right leads, 
\begin{equation}
H_{{\rm link}} = -t^{\prime }\sum_{\sigma }(a_{-1,\sigma }^{\dag }c_{0_1,\sigma
}+a_{1,\sigma }^{\dag }c_{L/2_1,\sigma }+{\rm H.c.})  \label{ehl2}
\end{equation}
describes the exchange of quasiparticles between the leads ($a_{i,\sigma }$)
and particular sites of leg 1 ($c_{i_1,\sigma }$), and 
\begin{eqnarray}
H_{{\rm ring}} &=&-e V_g \sum_{i,l,\sigma} 
c_{i_l,\sigma }^{\dag }c_{i_l,\sigma} -t_{\parallel}(c_{i_l,\sigma }^{\dag 
}c_{i_l+1,\sigma}e^{-i\phi/L} + {\rm H.c.}) \nonumber\\
&-&t_{\perp} \sum_{i,\sigma}(c_{i_1,\sigma }^{\dag }c_{i_2,\sigma } +{\rm 
H.c.})
+H_{{\rm int}}  \label{ehr}
\end{eqnarray}
describes the interacting electron system. The fermionic operators $c_{i_l,\sigma
}^{\dag }$ create an electron at site $i=1,L$ of leg $l=1,2$ with spin $\sigma$. The AB
ring has $L$ rungs, is threaded by a flux $\phi$ ($\phi = 2\pi\Phi/\Phi_0$), and is 
subjected to an applied gate voltage $V_g$.

\begin{figure}
\includegraphics[width=7.5cm]{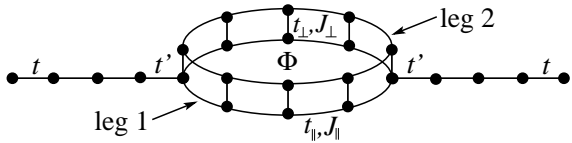}
\caption{Schematic representation of the 
system.}
\label{ladd}
\end{figure}

As we will consider the $t$-$J$ model for the ladder, the interacting part of the 
Hamiltonian reads:
\begin{equation}
H_{{\rm int}} = J_{\parallel}\sum_{i,l}{\bf S}_{i_l} {\bf \cdot 
S}_{i_l+1} 
+ J_{\perp}\sum_{i}{\bf S}_{i_1} {\bf \cdot S}_{i_2}
\end{equation}
where ${\bf S}_{i_l} = \sum_{\alpha\beta}c_{i_l \alpha}^{\dag} \sigma_{\alpha\beta} 
c_{i_l \beta}$ is the spin at site $i$ and leg $l$ and no double occupancy is 
allowed.

When the ground state is non-degenerate, the zero temperature transmission
from left to right can be
calculated to second order in $t^{\prime }$ by means of the retarded Green
function for the isolated ladder ring between sites $i$ and $j$: 
$G_{i,j}^{\mathrm{R%
}}(\omega)$, for an incident particle with energy $\omega$ and momentum $\pm
k$:~\cite{Jagla,ihm}
\begin{equation}
T(\omega,V_g,\phi) = \frac{4t^{2} \sin ^{2}k|{\tilde{t}}(\omega )|^{2}}{%
|[\omega - {\epsilon}(\omega) + te^{ik}]^{2}-|{\tilde{t}}^{2}(\omega )||^{2}}%
,  \label{tra}
\end{equation}
where
$\epsilon(\omega) = t^{\prime\,2} G_{0_1,0_1}^{\mathrm{R}}(\omega)$,
the effective hopping across the ring is
$\tilde{t}(\omega) = t^{\prime
\,2}G_{0_1,L/2_1}^{\mathrm{R}}(\omega )$, and $\omega = -2t
\cos k$ is the tight-binding dispersion relation for free electrons in
the leads. This equation is in fact exact for a non-interacting system; with
interactions on the ring it
serves as an approximation in the tunneling limit $t^{\prime}/t \ll 1$~\cite{Jagla,ihm} 
for a non-degenerate ground state. For an odd number of electrons,
the ground state
is Kramers degenerate for a system with time reversal symmetry and the equation
ceases to be valid.~\cite{ihm,lobos} So we assume that 
the ensuing Kondo effect is destroyed by temperature or magnetic field.~\cite{nos1,ihm}  
The conductance is $G=(2e^2/h)T$.

We calculate $T(\omega,V_g,\phi)$ by numerically diagonalizing the isolated
interacting ring in the presence of a magnetic field with $L$ rungs and $N$
electrons in the ground state to obtain the Green functions which appear in 
Eq.(\ref{tra}), fixing the
chemical potential of the non-interacting leads to zero ($\omega = 0$). By varying
$V_g$, $T(0,V_g,\phi)$ presents narrow peaks with a width proportional to
$(t^{\prime })^{2}$ at gate voltages which correspond to the excitation energies of
the system with $N-1$ particles.~\cite{nos1,ihm} The transmittance is 
obtained by 
integrating the spectra 
over a small energy window at the Fermi energy.~\cite{Jagla,meden,nos1}

\section{Results}

In order to study the robustness of the spin-charge separation in the presence of a 
second chain, we first show
results for weakly coupled chains ($t_{\perp} \ll t_{\parallel}$) and 
$J_i=0$ $(i = \perp, \parallel)$,
for which we know that in one chain there is complete SCS.~\cite{ogata,caspers,nos1}
In Fig. 2 we show the results for several small values of $t_{\perp}$ and 
in fact, observe clear dips at certain fractional values of the magnetic 
flux (the abrupt jumps correspond to other level crossings).
This is the first evidence, up to our knowledge of charge-spin 
separation in finite systems with more than one chain.

\begin{figure}
\includegraphics[width=7.5cm]{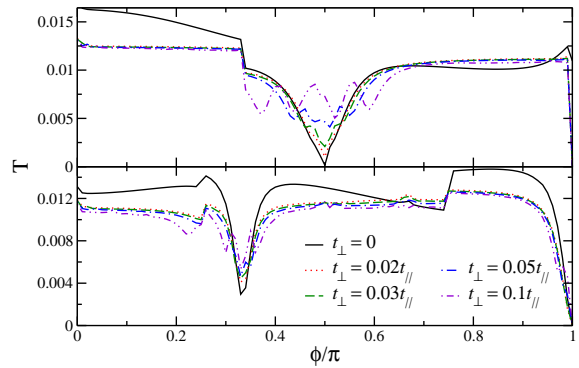}
\caption{
Weakly coupled chains: Transmittance as a function of flux for the 
anisotropic $t$-$J$ model with $J=0$, $t_{\parallel}=0.1$, $L=6$ rungs, 
$t^{\prime}=0.05 t_{\parallel}$, several values of
$t_{\perp}$ and $N=6$ (top) and $N=8$ (bottom) electrons in the ground 
state.
}
\label{fig1}
\end{figure}

To understand the position of the dips we resort to the expression obtained in 1D for 
$J=0$ $(i = \perp, \parallel)$ \cite{nos1,rincon2}. Considering a 
non-degenerate ground state containing 
$N=N_e+1$ particles and analyzing the part of the Green function that enters 
the 
transmittance when a particle is destroyed, it is shown that the dips occur when two 
intermediate states cross at a given flux and interfere destructively. These particular 
fluxes depend on the spin quantum numbers, and are located at
\begin{equation}
\phi _{d}=\pi (2n+1)/N_e,  
\label{dip}
\end{equation}
with $n$ integer.
If the integration energy window 
includes these levels, a dip in the conductance arises. 

For the ladder with $t_{\perp}=0$ and a total even number of electrons $N$ in the 
ground state, the
lowest-lying state has $N/2$ electrons in each leg. As we are calculating the 
transmittance through one leg only, and the intermediate state has one particle less, 
from the condition for $\phi_{d}$ with $N_e=N/2-1$, one expects
to see dips at 
$\phi_{d} = \pi \frac{2n+1}{N/2-1}$. In Fig. 2 we see that this is the case 
since for the top figure there will be $N_e+1=3$ electrons in each leg, leading to a dip 
at $\phi=\pi/2$ and for the bottom figure there will be 4 electrons in each leg 
leading to a dip at  $\phi=\pi/3$. 
When the ``second" dimension is turned on and  $t_{\perp}\neq 0$, we find that the 
dips remain and are quite robust, even for values of 
$t_{\perp}/t_{\parallel}$ as high as 0.1.

These are so far the results for a weak interchain coupling. Another 
atracting case is the one with a large coupling between the legs, 
$t_{\perp} \gg t_{\parallel}$. In this
limit and for the noninteracting case, the bands corresponding to the bonding and
antibonding states of each rung are very far apart and one might expect the reappearance
of SCS. In fact, this is the case, as can be seen in Fig. 3, where we plot the
transmittance for a ladder with several values of $t_{\perp}$ and fillings. Now the
total number of electrons in the lower band corresponds to the total filling $N$ (for a
less than half filled band) and the transmittance will involve $N_e=N-1$ electrons.
Hence, if SCS exists, the dips will be found at fluxes $\phi_{d} = \pi
\frac{2n+1}{N-1}$. In this figure we find that for large values of
$t_{\perp}/t_{\parallel}$, the dips correspond indeed to these fluxes. For smaller
values of $t_{\perp}$, we find a shift in the location of the minima and sometimes a
splitting of the dips.

\begin{figure}
\includegraphics[width=7.5cm]{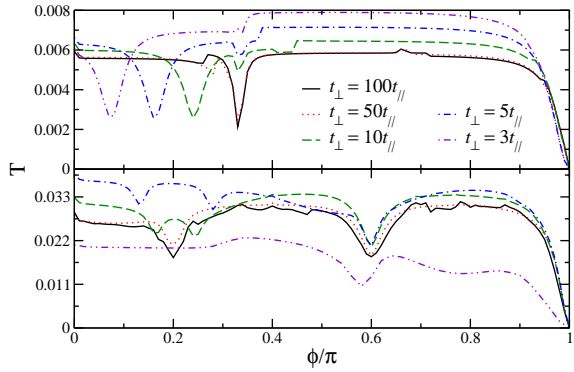}
\caption{
Strongly coupled chains: Flux-dependent transmittance for $L=6$ rungs,    
$J=0$, $t_{\parallel}=0.1$. Top: 
$t^{\prime}=0.05t_{\parallel}$ and $N=4$ electrons. Bottom: 
$t^{\prime}=0.09t_{\parallel}$ and $N=6$ electrons.
}
\label{fig2}
\end{figure}

\begin{figure}[b]
\vspace{2mm}
\includegraphics[width=7.5cm]{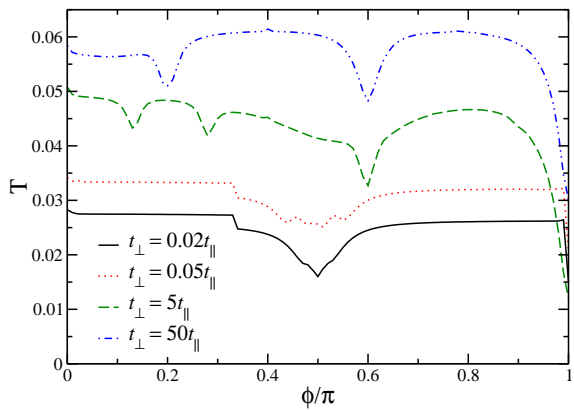}
\caption{Transition from weakly-coupled to strongly-coupled chains: 
Flux-dependent transmittance for several relations of inter to intrachain hoppings 
$t_{\perp}/t_{\parallel}$ for $L=6$ rungs and $N=6$ particles. 
}
\label{fig:allcouplings}
\end{figure}

It is interesting to visualize the behaviour of the dips when changing the parameters 
from weakly interacting chains to the strong coupling case 
(small to big $t_{\perp}/t_{\parallel}$). In Fig.\ref{fig:allcouplings} we collect the 
data for $N=6$ particles, where we see that for $t_{\perp}/t_{\parallel}\ll 1$ only 
one dip shows up (reflecting the bahaviour of half the number of particles in each band 
corresponding mainly to each leg of the ladder as in Fig. 2). On the contrary, for  
$t_{\perp}/t_{\parallel}\gg 1$ all $N$ particles belong to the relevant lower (bonding) 
band and two main dips arise at the positions corresponding to this strongly coupled 
case as in Fig. 3. 

\section{Effective model for strongly coupled chains}

In order to study the strongly coupled case and understand the transition towards
the limit of isotropic exchange, we have mapped our ladder Hamiltonian $H_{{\rm ring}}$ 
to an effective model in the subspace of the bonding states of each rung,
using degenerate perturbation theory up to second order in $t_{\parallel }$.~\cite{pt}
The model is valid for energies lower than $t_{\perp }$
and a less than half-filled system:
\begin{eqnarray}
H_{{\rm eff}} &=&-t_{\parallel }\sum_{i\sigma }(\hat{c}_{i\sigma }^{\dag 
}\hat{c}_{i+1,\sigma 
}+\mathrm{H.c.})+J\sum_{i}(\mathbf{S}_{i}\mathbf{\cdot S}_{i+1}-1/4)
\nonumber \\
&&+t^{\prime \prime }\sum_{i\sigma }(\hat{c}_{i+2,\sigma }^{\dag }\hat{c}_{i\sigma }(%
\mathbf{S}_{i}\mathbf{\cdot S}_{i+1}-1/4)+\mathrm{H.c.}),  \label{heff}
\end{eqnarray}%
where $J=J_{\parallel }/2+2t_{\parallel }^{2}/(t_{\perp }-3J_{\perp }/4)$, $t^{\prime 
\prime }=t_{\parallel }^{2}/(t_{\perp }-3J_{\perp }/4)$ and 
$\hat{c}_{i\sigma}=\frac{1}{\sqrt{2}}(c_{i_1,\sigma }+c_{i_2,\sigma })$, the bonding 
operator.

The second dimension of the original ladder is reflected by the second 
nearest-neighbour term which tends to destroy the dips. 
In Fig. 5 we show the transmittance of a particle through a 
bonding channel and compare it with that of $H_{{\rm eff}}$, finding excellent
agreement 
for the strongly coupled case. As $t_{\perp}/t_{\parallel}$ diminishes, the curves start 
to differ as $H_{{\rm eff}}$ loses its validity. Comparing with Fig. 3, where a 
particle is injected to one site only instead of into a bonding state, we find that 
the effective model is also valid for 
this case, as long as $t_{\perp} \gg t_{\parallel}$, since the antibonding band is 
shifted to upper energies.

\begin{figure}
\includegraphics[width=7.5cm]{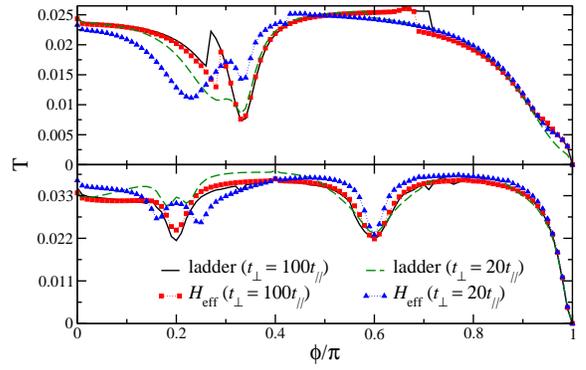}
\caption{
Transmittance through a bonding channel and comparison with $H_{{\rm eff}}$ for the 
same system as in Fig. 3.}
\label{fig3}
\end{figure}

\begin{figure}
\vspace{4mm}
\includegraphics[width=7.5cm]{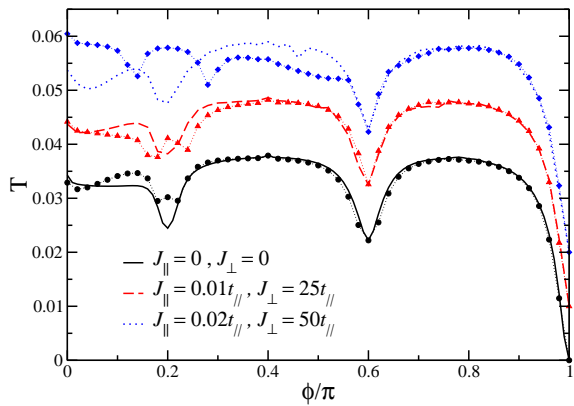}
\caption{
Transmittance through a bonding channel (lines) and comparison with $H_{{\rm eff}}$ 
(symbols) for finite interactions $J_{\parallel}$ and $J_{\perp}$ 
for $t_{\perp}=50t_{\parallel}$, $L=6$ rungs and $N=6$ electrons.
}
\label{fig:jotafinito}
\end{figure}

So far we have presented results for $J_{\parallel}=J_{\perp}=0$. Finite interactions  
introduce an extra spin shuffling 
in the system, mixing the spin wave quantum numbers. In spite of the fact that the 
conditions that lead to Eq.(5) (based on the $J=0$ limit) are no longer valid, dips are 
still observed for small values of the interactions in the strongly coupled case.
However, the effect of the interaction is to 
reduce the depth of the dips and shift their position. In Fig.\ref{fig:jotafinito} we 
show results for the 6-rung ladder with $N=6$ particles in the ground state and several 
values of the interactions. Taking into account the fact that the $J$'s are obtained 
perturbatively from the large-$U$ Hubbard model, we keep their relation as 
 $J_{\perp}/J_{\parallel}=t_{\perp}^2/t_{\parallel}^2$. Two observations can be made: 
i) The dips are still present for finite $J$'s. ii) However, as for the $J=0$ case where 
the dips were affected by the interchain hopping parameter, in this case we also find 
shifts and reductions in the their depth caused by the interactions. A similar behaviour 
occurs for weakly interacting chains. We also find that the effective model fits 
quite well the results in the ladder for finite $J_\perp$ and 
$J_\parallel$, and its range of 
validity extends to appreciable values of the interaction parameters.

\section{Conclusions}

In summary, we have found that the dips in the conductance, predicted to appear in
strongly correlated chains as a consequence of spin-charge separation, are robust in
the presence of a second transmission channel modelled by a ladder system in the
anisotropic limit. For a wide range of parameters, in particular for weak and strong
hopping couplings across the rungs $t_{\perp }$, the dips remain. 
However, their position 
differ from the predictions stemming from the exactly solvable case of the 
Hubbard chain with infinite $U$ (or $t$-$J$ model with $J=0$).~\cite{nos1} 
For intermediate values of
$t_{\perp }$ the dips disappear.  The signatures of spin-charge separation are also 
robust for finite, albeit small, values of spin-spin interactions.
This is the first time in which 
signatures of spin-charge separation are observed 
in interacting finite systems with more than one chain, thus 
opening the possibility of measuring this peculiar phenomenon in real nanoscopic systems.

It is a pleasure to thank S. Ramasesha and M. Kumar for helpful
discussions. The authors are supported by CONICET. This work was done in the 
framework of projects PIP 5254 of CONICET and PICT 2006/483 of the ANPCyT.

\end{document}